\documentclass[12pt]{iopart}

\usepackage{xspace}
\usepackage{iopams}  
\usepackage{amsmath}
\usepackage{color}
\usepackage{graphicx}
\usepackage{appendix}
\usepackage{caption}
\usepackage{acronym}
\usepackage{orcidlink}
\usepackage{lipsum}

\acrodef{LVK}[LVK]{LIGO, Virgo and KAGRA}
\acrodef{LDG}[LDG]{LIGO Data Grid}

\acrodef{L1}[L]{LIGO Livingston}
\acrodef{H1}[H]{LIGO Hanford}
\acrodef{V1}[V]{Virgo}

\acrodef{O4}[O4]{fourth observing run}
\acrodef{O3}[O3]{third observing run}
\acrodef{O2}[O2]{second observing run}
\acrodef{O1}[O1]{first observing run}

\acrodef{GWTC}[GWTC-3]{Gravitational Wave Transient Catalog}
\acrodef{EM}[EM]{electromagnetic}
\acrodef{GRB}[GRB]{gamma-ray burst}

\acrodef{GPS}[GPS]{Global Positioning System}
\acrodef{GCN}[GCN]{Gamma-ray Coordinate Network}

\acrodef{BH}[BH]{black hole}
\acrodef{IMBH}[IMBH]{intermediate mass black hole}
\acrodef{BBH}[BBH]{binary black hole}
\acrodefplural{BBH}[BBHs]{binary black holes}
\acrodef{BNS}[BNS]{binary neutron star}
\acrodef{NS}[NS]{neutron star}
\acrodef{NSBH}[NSBH]{neutron star--black hole binary}
\acrodefplural{NSBH}[NSBHs]{neutron star--black hole binaries}
\acrodef{CBC}[CBC]{compact binary coalescence}
\acrodefplural{CBC}[CBCs]{compact binary coalescences}
\acrodef{GW}[GW]{gravitational wave}
\acrodef{SSM}[SSM]{sub-solar mass black hole}

\acrodef{SNR}[SNR]{signal-to-noise ratio}
\acrodefplural{SNR}[SNRs]{signal-to-noise ratios}
\acrodef{FAR}[FAR]{false alarm rate}
\acrodefplural{FAR}[FARs]{false alarm rates}
\acrodef{IFAR}[IFAR]{inverse false alarm rate}
\acrodef{PSD}[PSD]{power spectral density}
\acrodefplural{PSD}[PSDs]{power spectral densities}
\acrodef{SVD}[SVD]{singular value decomposition}
\acrodef{FIR}[FIR]{finite impulse response}
\acrodef{MDC}[MDC]{Mock Data Challenge}
\acrodef{LLOID}[LLOID]{Low-Latency Online Inspiral Detection}
\acrodef{PN}[PN]{Post-Newtonian}
\acrodef{FFT}[FFT]{fast Fourier transform}

\acrodef{GRACEDB}[GraceDB]{Gravitational wave Candidate Event Database}
\acrodef{GWLTS}[\texttt{gw-lts}]{Gravitational Wave Low-latency Test Suite}
\acrodef{IDQ}[iDQ]{integrated Data Quality}

\newcommand\hmm[1]{\ifnum\ifhmode\spacefactor\else2000\fi>1500 \uppercase{#1}\else#1\fi}

\newcommand{\GSTLAL}{GstLAL\xspace}

\newcommand{\MDCSTART}{5 Jan. 2020 15:59:42}
\newcommand{\MDCEND}{14 Feb. 2020 15:59:42}

\newcommand{\PASTRO}{\ensuremath{\mathrm{P}_{\mathrm{astro}}}}

\newcommand{\ALLINBANKEFFICIENCY}[1]{%
	\IfEqCase{#1}{%
		{ONEPERHOUR}{\ensuremath{0.87}}%
		{TWOPERDAY}{\ensuremath{0.84}}%
		{ONEPERMONTH}{\ensuremath{0.77}}%
		{TWOPERYEAR}{\ensuremath{0.72}}%
	}[\PackageError{ALLINBANKEFFICIENCY}{Undefined option: #1}{}]
}%

\newcommand{\BBHINBANKEFFICIENCY}[1]{%
	\IfEqCase{#1}{%
		{ONEPERHOUR}{\ensuremath{0.87}}%
		{TWOPERDAY}{\ensuremath{0.84}}%
		{ONEPERMONTH}{\ensuremath{0.75}}%
		{TWOPERYEAR}{\ensuremath{0.69}}%
	}[\PackageError{BBHINBANKEFFICIENCY}{Undefined option: #1}{}]
}%

\newcommand{\BNSINBANKEFFICIENCY}[1]{%
	\IfEqCase{#1}{%
		{ONEPERHOUR}{\ensuremath{0.95}}%
		{TWOPERDAY}{\ensuremath{0.94}}%
		{ONEPERMONTH}{\ensuremath{0.89}}%
		{TWOPERYEAR}{\ensuremath{0.86}}%
	}[\PackageError{BNSINBANKEFFICIENCY}{Undefined option: #1}{}]
}%

\newcommand{\NSBHINBANKEFFICIENCY}[1]{%
	\IfEqCase{#1}{%
		{ONEPERHOUR}{\ensuremath{0.77}}%
		{TWOPERDAY}{\ensuremath{0.71}}%
		{ONEPERMONTH}{\ensuremath{0.65}}%
		{TWOPERYEAR}{\ensuremath{0.62}}%
	}[\PackageError{NSBHINBANKEFFICIENCY}{Undefined option: #1}{}]
}%

\newcommand{\MEAN}[1]{%
	\IfEqCase{#1}{%
		{MASSRATIO}{\ensuremath{1.39}}%
		{MCHIRP}{\ensuremath{0.15}}%
		{SPIN1Z}{\ensuremath{7.27}}%
		{SPIN2Z}{\ensuremath{2.81}}%
		{CHIEFF}{\ensuremath{5.77}}%
		{ENDTIME}{\ensuremath{6.17}}%
	}[\PackageError{MEAN}{Undefined option: #1}{}]
}%

\newcommand{\STDEV}[1]{%
	\IfEqCase{#1}{%
		{MASSRATIO}{\ensuremath{2.86}}%
		{MCHIRP}{\ensuremath{0.45}}%
		{SPIN1Z}{\ensuremath{285}}%
		{SPIN2Z}{\ensuremath{183}}%
		{CHIEFF}{\ensuremath{252}}%
		{ENDTIME}{\ensuremath{33.8}}%
	}[\PackageError{STDEV}{Undefined option: #1}{}]
}%

\newcommand{\QFIFTY}[1]{%
	\IfEqCase{#1}{%
		{MASSRATIO}{\ensuremath{0.45}}%
		{MCHIRP}{\ensuremath{0.007}}%
		{SPIN1Z}{\ensuremath{1.24}}%
		{SPIN2Z}{\ensuremath{1.72}}%
		{CHIEFF}{\ensuremath{1.34}}%
		{ENDTIME}{\ensuremath{15.9}}%
	}[\PackageError{QFIFTY}{Undefined option: #1}{}]
}%

\newcommand{\QSEVENTYFIVE}[1]{%
	\IfEqCase{#1}{%
		{MASSRATIO}{\ensuremath{1.67}}%
		{MCHIRP}{\ensuremath{0.33}}%
		{SPIN1Z}{\ensuremath{3.82}}%
		{SPIN2Z}{\ensuremath{5.33}}%
		{CHIEFF}{\ensuremath{3.71}}%
		{ENDTIME}{\ensuremath{20.4}}%
	}[\PackageError{QSEVENTYFIVE}{Undefined option: #1}{}]
}%

\newcommand{\QNINETY}[1]{%
	\IfEqCase{#1}{%
		{MASSRATIO}{\ensuremath{4.97}}%
		{MCHIRP}{\ensuremath{0.73}}%
		{SPIN1Z}{\ensuremath{13.8}}%
		{SPIN2Z}{\ensuremath{17.5}}%
		{CHIEFF}{\ensuremath{10.8}}%
		{ENDTIME}{\ensuremath{26.4}}%
	}[\PackageError{QNINETY}{Undefined option: #1}{}]
}%

\newcommand{\INJECTEDVT}[1]{%
	\IfEqCase{#1}{%
		{BNS}{\ensuremath{1.08\times10^{-1}}}%
		{NSBH}{\ensuremath{4.34\times10^{-1}}}%
		{BBH}{\ensuremath{29.1}}%
	}[\PackageError{INJECTEDVT}{Undefined option: #1}{}]
}%

\newcommand{\VTTWOPERDAY}[1]{%
	\IfEqCase{#1}{%
		{BNS}{\ensuremath{3.49\times10^{-4}}}%
		{NSBH}{\ensuremath{8.08\times10^{-4}}}%
		{BBH}{\ensuremath{1.23\times10^{-1}}}%
	}[\PackageError{VTTWOPERDAY}{Undefined option: #1}{}]
}%

\newcommand{\VTNETSNR}[1]{%
	\IfEqCase{#1}{%
		{BNS}{\ensuremath{4.41\times10^{-4}}}%
		{NSBH}{\ensuremath{1.59\times10^{-3}}}%
		{BBH}{\ensuremath{1.52\times10^{-1}}}%
	}[\PackageError{VTNETSNR}{Undefined option: #1}{}]
}%

\newcommand{\VTDECSNR}[1]{%
	\IfEqCase{#1}{%
		{BNS}{\ensuremath{1.47\times10^{-4}}}%
		{NSBH}{\ensuremath{4.98\times10^{-3}}}%
		{BBH}{\ensuremath{5.46\times10^{-2}}}%
	}[\PackageError{VTDECSNR}{Undefined option: #1}{}]
}%

\newcommand{\SEARCHEDAREAQFIFTY}[1]{%
	\IfEqCase{#1}{%
		{ALL}{\ensuremath{271}}%
		{TRIPLE}{\ensuremath{31.9}}%
		{DOUBLE}{\ensuremath{301}}%
		{SINGLE}{\ensuremath{3150}}%
	}[\PackageError{SEARCHEDAREAQFIFTY}{Undefined option: #1}{}]
}%

\newcommand{\SEARCHEDAREAQSEVENTYFIVE}[1]{%
	\IfEqCase{#1}{%
		{ALL}{\ensuremath{1080}}%
		{TRIPLE}{\ensuremath{140}}%
		{DOUBLE}{\ensuremath{893}}%
		{SINGLE}{\ensuremath{10,400}}%
	}[\PackageError{SEARCHEDAREAQSEVENTYFIVE}{Undefined option: #1}{}]
}%

\newcommand{\SEARCHEDAREAQNINETY}[1]{%
	\IfEqCase{#1}{%
		{ALL}{\ensuremath{3910}}%
		{TRIPLE}{\ensuremath{357}}%
		{DOUBLE}{\ensuremath{2470}}%
		{SINGLE}{\ensuremath{18,400}}%
	}[\PackageError{SEARCHEDAREAQNINETY}{Undefined option: #1}{}]
}%

\newcommand{\SEARCHEDPROBQFIFTY}[1]{%
	\IfEqCase{#1}{%
		{ALL}{\ensuremath{0.53}}%
		{TRIPLE}{\ensuremath{0.58}}%
		{DOUBLE}{\ensuremath{0.52}}%
		{SINGLE}{\ensuremath{0.59}}%
	}[\PackageError{SEARCHEDPROBQFIFTY}{Undefined option: #1}{}]
}%

\newcommand{\SEARCHEDPROBQSEVENTYFIVE}[1]{%
	\IfEqCase{#1}{%
		{ALL}{\ensuremath{0.79}}%
		{TRIPLE}{\ensuremath{0.84}}%
		{DOUBLE}{\ensuremath{0.77}}%
		{SINGLE}{\ensuremath{0.78}}%
	}[\PackageError{SEARCHEDPROBQSEVENTYFIVE}{Undefined option: #1}{}]
}%

\newcommand{\SEARCHEDPROBQNINETY}[1]{%
	\IfEqCase{#1}{%
		{ALL}{\ensuremath{0.93}}%
		{TRIPLE}{\ensuremath{0.96}}%
		{DOUBLE}{\ensuremath{0.92}}%
		{SINGLE}{\ensuremath{0.92}}%
	}[\PackageError{SEARCHEDPROBQNINETY}{Undefined option: #1}{}]
}%

\newcommand{\OTHREEOPA}{\ensuremath{1.2}} 

\newcommand{\MDCGWIFOS}[1]{%
	\IfEqCase{#1}{%
		{GW200112}{L1}%
		{GW200115}{H1L1}%
		{GW200128}{H1L1}%
		{GW200129}{H1L1V1}%
		{GW200202}{H1L1}%
		{GW200208q}{H1L1}%
		{GW200208am}{H1L1}%
		{GW200209}{H1L1}%
		{GW200210}{H1L1}%
	}[\PackageError{MDCGWIFOS}{Undefined option: #1}{}]
}%

\newcommand{\MDCGWSNR}[1]{%
	\IfEqCase{#1}{%
		{GW200112}{\ensuremath{18.46}}%
		{GW200115}{\ensuremath{11.48}}%
		{GW200128}{\ensuremath{9.98}}%
		{GW200129}{\ensuremath{26.30}}%
		{GW200202}{\ensuremath{11.09}}%
		{GW200208q}{\ensuremath{10.56}}%
		{GW200208am}{\ensuremath{8.00}}%
		{GW200209}{\ensuremath{9.96}}%
		{GW200210}{\ensuremath{9.28}}%
	}[\PackageError{MDCGWSNR}{Undefined option: #1}{}]
}%

\newcommand{\MDCGWFAR}[1]{%
	\IfEqCase{#1}{%
		{GW200112}{\ensuremath{1.01\times10^{-7}}}%
		{GW200115}{\ensuremath{2.55\times10^{-4}}}%
		{GW200128}{\ensuremath{1.44\times10^{-4}}}%
		{GW200129}{\ensuremath{1.78\times10^{-17}}}%
		{GW200202}{\ensuremath{1.69\times10^{-2}}}%
		{GW200208q}{\ensuremath{4.92\times10^{-5}}}%
		{GW200208am}{\ensuremath{2.02\times10^{3}}}%
		{GW200209}{\ensuremath{1.20}}%
		{GW200210}{\ensuremath{3.64\times10^{3}}}%
	}[\PackageError{MDCGWFAR}{Undefined option: #1}{}]
}%

\newcommand{\MDCGWPASTRO}[1]{%
	\IfEqCase{#1}{%
		{GW200112}{\ensuremath{>0.99}}%
		{GW200115}{\ensuremath{>0.99}}%
		{GW200128}{\ensuremath{>0.99}}%
		{GW200129}{\ensuremath{>0.99}}%
		{GW200202}{\ensuremath{>0.99}}%
		{GW200208q}{\ensuremath{>0.99}}%
		{GW200208am}{\ensuremath{0.48}}%
		{GW200209}{\ensuremath{>0.99}}%
		{GW200210}{0.27}%
	}[\PackageError{MDCGWPASTRO}{Undefined option: #1}{}]
}%

\newcommand{\MDCGWMCHIRP}[1]{%
	\IfEqCase{#1}{%
		{GW200112}{\ensuremath{33.37~M_{\odot}}}%
		{GW200115}{\ensuremath{2.58~M_{\odot}}}%
		{GW200128}{\ensuremath{50.74~M_{\odot}}}%
		{GW200129}{\ensuremath{30.66~M_{\odot}}}%
		{GW200202}{\ensuremath{8.15~M_{\odot}}}%
		{GW200208q}{\ensuremath{34.50~M_{\odot}}}%
		{GW200208am}{\ensuremath{66.59~M_{\odot}}}%
		{GW200209}{\ensuremath{39.45~M_{\odot}}}%
		{GW200210}{\ensuremath{7.89~M_{\odot}}}%
	}[\PackageError{MDCGWMCHIRP}{Undefined option: #1}{}]
}%

\newcommand{\OTHREEGWIFOS}[1]{%
	\IfEqCase{#1}{%
		{GW200112}{L1}%
		{GW200115}{H1L1}%
		{GW200128}{--}%
		{GW200129}{H1L1V1}%
		{GW200202}{--}%
		{GW200208q}{--}%
		{GW200208am}{--}%
		{GW200209}{--}%
		{GW200210}{--}%
	}[\PackageError{OTHREEGWIFOS}{Undefined option: #1}{}]
}%

\newcommand{\OTHREEGWSNR}[1]{%
	\IfEqCase{#1}{%
		{GW200112}{\ensuremath{18.79}}%
		{GW200115}{\ensuremath{11.42}}%
		{GW200128}{--}%
		{GW200129}{\ensuremath{26.61}}%
		{GW200202}{--}%
		{GW200208q}{--}%
		{GW200208am}{--}%
		{GW200209}{--}%
		{GW200210}{--}%
	}[\PackageError{OTHREEGWSNR}{Undefined option: #1}{}]
}%

\newcommand{\OTHREEGWFAR}[1]{%
	\IfEqCase{#1}{%
		{GW200112}{\ensuremath{4.05\times10^{-4}}}%
		{GW200115}{\ensuremath{6.61\times10^{-4}}}%
		{GW200128}{\ensuremath{> \OTHREEOPA{}}}%
		{GW200129}{\ensuremath{2.11\times10^{-24}}}%
		{GW200202}{\ensuremath{> \OTHREEOPA{}}}%
		{GW200208q}{--}%
		{GW200208am}{--}%
		{GW200209}{--}%
		{GW200210}{--}%
	}[\PackageError{OTHREEGWFAR}{Undefined option: #1}{}]
}%

\newcommand{\OTHREEGWPASTRO}[1]{%
	\IfEqCase{#1}{%
		{GW200112}{\ensuremath{>0.99}}%
		{GW200115}{\ensuremath{>0.99}}%
		{GW200128}{--}%
		{GW200129}{\ensuremath{>0.99}}%
		{GW200202}{--}%
		{GW200208q}{--}%
		{GW200208am}{--}%
		{GW200209}{--}%
		{GW200210}{--}%
	}[\PackageError{OTHREEGWPASTRO}{Undefined option: #1}{}]
}%

\newcommand{\OTHREEGWMCHIRP}[1]{%
	\IfEqCase{#1}{%
		{GW200112}{\ensuremath{35.37~M_{\odot}}}%
		{GW200115}{\ensuremath{2.57~M_{\odot}}}%
		{GW200128}{--}%
		{GW200129}{\ensuremath{32.74~M_{\odot}}}%
		{GW200202}{--}%
		{GW200208q}{--}%
		{GW200208am}{--}%
		{GW200209}{--}%
		{GW200210}{--}%
	}[\PackageError{OTHREEGWMCHIRP}{Undefined option: #1}{}]
}

\usepackage[minnames=4,maxnames=4,style=numeric,backend=bibtex]{biblatex}
\bibliography{references}
\graphicspath{{figures/}}
\usepackage{float}
\begin{document}

\title[]{When to Point Your Telescopes: Gravitational Wave Trigger Classification for Real-Time Multi-Messenger Followup Observations}

\author{Anarya Ray \orcidlink{0000-0002-7322-4748}$^{1,*}$, Wanting Niu \orcidlink{0000-0003-1470-532X}$^{2,3}$, Shio Sakon \orcidlink{0000-0002-5861-3024}$^{2,3}$, Becca Ewing$^{2,3}$, Jolien D. E. Creighton \orcidlink{0000-0003-3600-2406}$^{1}$, Chad Hanna$^{2,3}$, Shomik Adhicary$^{2,3}$, Pratyusava Baral \orcidlink{0000-0001-6308-211X}$^{1}$, Amanda Baylor \orcidlink{0000-0003-0918-0864}$^{1}$, Kipp Cannon \orcidlink{0000-0003-4068-6572}$^{4}$, Sarah Caudill$^{5,6}$, Bryce Cousins \orcidlink{0000-0002-7026-1340}$^{7,2,3}$, Heather Fong$^{8,4,9}$, Richard N. George \orcidlink{0000-0002-7797-7683}$^{10}$, Patrick Godwin$^{11,2,3}$, Reiko Harada$^{4,9}$, Yun-Jing Huang \orcidlink{0000-0002-2952-8429}$^{2,3}$, Rachael Huxford$^{2,3}$, Prathamesh Joshi \orcidlink{0000-0002-4148-4932}$^{2,3}$, Shasvath Kapadia \orcidlink{0000-0001-5318-1253}$^{12}$, James Kennington \orcidlink{0000-0002-6899-3833}$^{2,3}$, Soichiro Kuwahara$^{4,9}$, Alvin K. Y. Li \orcidlink{0000-0001-6728-6523}$^{11}$, Ryan Magee \orcidlink{0000-0001-9769-531X}$^{11}$, Duncan Meacher \orcidlink{0000-0001-5882-0368}$^{1}$, Cody Messick$^{13}$, Soichiro Morisaki \orcidlink{0000-0002-8445-6747}$^{14,1}$, Debnandini Mukherjee  \orcidlink{0000-0001-7335-9418}$^{15,16}$, Alex Pace$^{2,3}$, Cort Posnansky$^{2,3}$, Surabhi Sachdev \orcidlink{0000-0002-0525-2317}$^{17,1}$, Divya Singh \orcidlink{0000-0001-9675-4584}$^{2,3}$, Ron Tapia$^{2,18}$, Leo Tsukada  \orcidlink{0000-0003-0596-5648}$^{2,3}$, Takuya Tsutsui \orcidlink{0000-0002-2909-0471}$^{4}$, Koh Ueno \orcidlink{0000-0003-3227-6055}$^{4}$, Aaron Viets \orcidlink{0000-0002-4241-1428}$^{19}$, Leslie Wade$^{20}$, Madeline Wade \orcidlink{0000-0002-5703-4469}$^{20}$} 

\address{$^{1}$ Leonard E.\ Parker Center for Gravitation, Cosmology, and Astrophysics, University of Wisconsin-Milwaukee, Milwaukee, WI 53201, USA}
\address{$^{2}$Department of Physics, The Pennsylvania State University, University Park, PA 16802, USA}
\address{$^{3}$Institute for Gravitation and the Cosmos, The Pennsylvania State University, University Park, PA 16802, USA}
\address{$^{4}$RESCEU, The University of Tokyo, Tokyo, 113-0033, Japan}
\address{$^{5}$Department of Physics, University of Massachusetts, Dartmouth, MA 02747, USA}
\address{$^{6}$Center for Scientific Computing and Data Science Research, University of Massachusetts, Dartmouth, MA 02747, USA}
\address{$^{7}$Department of Physics, University of Illinois, Urbana, IL 61801 USA}
\address{$^{8}$Department of Physics and Astronomy, University of British Columbia, Vancouver, BC, V6T 1Z4, Canada}
\address{$^{9}$Graduate School of Science, The University of Tokyo, Tokyo 113-0033, Japan}
\address{$^{10}$Center for Gravitational Physics, University of Texas at Austin, Austin, TX 78712, USA}
\address{$^{11}$LIGO Laboratory, California Institute of Technology, MS 100-36, Pasadena, California 91125, USA}
\address{$^{12}$The Inter-University Centre for Astronomy and Astrophysics, Post Bag 4, Ganeshkhind, Pune 411007, India}
\address{$^{13}$MIT Kavli Institute for Astrophysics and Space Research, Massachusetts Institute of Technology, Cambridge, MA 02139, USA}
\address{$^{14}$Institute for Cosmic Ray Research, The University of Tokyo, 5-1-5 Kashiwanoha, Kashiwa, Chiba 277-8582, Japan}
\address{$^{15}$NASA Marshall Space Flight Center, Huntsville, AL 35811, USA}
\address{$^{16}$Center for Space Plasma and Aeronomic Research, University of Alabama in Huntsville, Huntsville, AL 35899, USA}
\address{$^{17}$School of Physics, Georgia Institute of Technology, Atlanta, GA 30332, USA}
\address{$^{18}$Institute for Computational and Data Sciences, The Pennsylvania State University, University Park, PA 16802, USA}
\address{$^{19}$Concordia University Wisconsin, Mequon, WI 53097, USA}
\address{$^{20}$Department of Physics, Hayes Hall, Kenyon College, Gambier, Ohio 43022, USA}
\ead{$^{*}$anarya@uwm.edu}
\vspace{10pt}
\begin{indented}
\item[]April 2023
\end{indented}

\begin{abstract}
We develop a robust and self-consistent framework to extract and classify gravitational wave candidates from noisy data, for the purpose of assisting in real-time multi-messenger follow-ups during LIGO-Virgo-KAGRA's fourth observing run~(O4). Our formalism implements several improvements to the low latency calculation of the probability of astrophysical origin~(\PASTRO{}), so as to correctly account for various factors such as the sensitivity change between observing runs, and the deviation of the recovered template waveform from the true gravitational wave signal that can strongly bias said calculation. We demonstrate the high accuracy with which our new formalism recovers and classifies gravitational wave triggers, by analyzing replay data from previous observing runs injected with simulated sources of different categories. We show that these improvements enable the correct identification of the majority of simulated sources, many of which would have otherwise been misclassified. We carry out the aforementioned analysis by implementing our formalism through the \GSTLAL{} search pipeline even though it can be used in conjunction with potentially any matched filtering pipeline. Armed with robust and self-consistent \PASTRO{} values, the \GSTLAL{} pipeline can be expected to provide accurate source classification information for assisting in multi-messenger follow-up observations to gravitational wave alerts sent out during O4. 
\end{abstract}

\section{Introduction}
The observation of \acp{GW}, a burst of gamma rays and other \ac{EM} transients in X-ray through radio, from the \ac{BNS} merger GW170817, marked a significant breakthrough in multi-messenger astronomy~\cite{170817GW,grb170817fermi,grb170817integral,GRBGW170817,SSS17a,170817multimessenger}. Interpretation of the complementary information carried by photons and \acp{GW} from a single \ac{BNS} system has provided new insights into several mysteries of fundamental physics, beyond the scope of what either messenger could have conveyed individually. Some examples include explorations of the properties of matter at extreme densities~\cite{multieos}, tests of general relativity in the strong gravity regime~\cite{multitgr,GRBGW170817}, the measurement of cosmological parameters~\cite{multicosmo}, and investigations on the production of heavy elements in the universe~\cite{multiheavy}.

To facilitate the prompt discovery of multi-messenger counterparts to GW observations, the \ac{LVK}~\cite{aligo,avirgo,kagra} detector network sends out public alerts about GW candidates, within minutes of their detection~\cite{observingscenarios}. These alerts comprise information regarding  source properties and candidate significance which can assist in optimizing the triggered follow-up campaigns. For example, the optical counterpart to GW170817 was detected while performing a targeted search that made use of the precise sky localization obtained through the analysis of multi-detector \ac{GW} data~\cite{SSS17a,170817multimessenger}.

For \acp{GW} originating from \acp{CBC}, information regarding the nature of the merging compact objects is also considered highly important for conducting real-time multi-messenger searches. The main reason for this consideration is as follows. Among the three kinds of \ac{CBC} sources expected to be observable by the \ac{LVK}, \acp{BNS} and \acp{NSBH} are far more likely to be associated with multi-messenger counterparts than \acp{BBH}~\cite{Metzger2019,Perna2019}. Hence, estimates of the probability that a CBC candidate has originated from either a \ac{BNS} or an \ac{NSBH}, that are sent out as part of \ac{GW} candidate alerts~\cite{observingscenarios}, can assist astronomers in making informed decisions on whether to follow-up on said alerts.

This so-called probability of astrophysical origin~(\PASTRO{}) can be computed from \ac{GW} data, given knowledge of the rate at which \ac{BBH}, \ac{BNS} and \ac{NSBH} systems merge in the observable universe and the population-level distributions of their system parameters~\cite{multifgmc,gwtc-1}. Informed by the relative rates and population properties of various \ac{CBC} sources, estimates of \PASTRO{} can be expected to reveal a larger number of significant candidates in densely populated regions of the \ac{CBC} parameter space, in complement to estimates of \ac{FAR}. Hence, in addition to facilitating source classification, the \PASTRO{} calculation also offers a population-informed measure of significance for \ac{GW} candidates. 

Several frameworks for estimating \PASTRO{} self-consistently with the rates and population properties of \acp{CBC} from \ac{GW} data have been studied in existing literature~\cite{fgmc,multifgmc,pop-pastro0,pop-pastro1}. However, they all involve constructing the posterior probability distributions of said rates given data associated with all the \ac{GW} candidates found above some \ac{FAR} threshold during an observing run. Since the mentioned dataset can only become available in postprocessing, such analyses cannot be used straightforwardly for computing \PASTRO{} in real-time, without resorting to various schemes for approximating the rate of detectable \acp{CBC}. 

In this paper, we describe a self-consistent methodology for computing \PASTRO{} in low-latency with the purpose of assisting in real-time multimessenger follow-up observations during \ac{LVK}'s \ac{O4}. Our framework comprises several improvements to the low-latency \PASTRO{} calculation that was implemented in \ac{LVK}'s \ac{O3}. We implement our formalism through the \GSTLAL{} search pipeline's real-time (\textit{online}) analysis of \ac{O4} data, even though our improved rate approximation scheme is applicable to other pipelines as well. 

We validate our formalism through \GSTLAL{}'s participation in a rigorous \ac{MDC} that comprised a replay of \ac{O3} data with simulated sources injected into the data stream~\cite{mdc_analytics,ewing2023performance}. We show that our formalism yields highly accurate category specific \PASTRO{} for both real and simulated events in the \ac{MDC}. We also demonstrate how our improved rate-approximation prevents the underestimation of \PASTRO{} values of real and simulated events that would have resulted from the use of \ac{O3}'s inaccurate rate-approximation scheme.

This paper is organized as follows. In section \ref{methods} we describe our formalism for computing low-latency \PASTRO{} values in detail. In section~\ref{sec:results}, we summarize the performance of our methods in the aforementioned MDC and demonstrate the accuracy of our category-specific \PASTRO{} values. In section~\ref{sec:conclusion}, we conclude with a summary of our methods, their validation, and future prospects regarding further improvements to our low-latency computation of \PASTRO{}.

\section{Methods}
\label{methods}
In this section, we describe our formalism for calculating accurate \PASTRO{} values in low latency amidst an ongoing observing run. We first summarize the offline calculation that can be implemented given the full set of GW triggers found during an observing run. We then derive several approximation schemes for accurate \PASTRO{} calculation in real-time, i.e. in the absence of the full list of triggers, which can only become available in post-processing. 
\subsection{Probability of astrophysical origin}
\label{methods:offline}
GW triggers originating from different sources (including terrestrial ones) can be interpreted as realizations of independent Poisson processes~\cite{fgmc,multifgmc}. Hence, given the rate of events expected from each source category and the GW data associated with a particular trigger, one can compute the probability of its origin being any one of said categories. For example, the probability of the $k^{th}$ trigger originating from category $\alpha$ is given by:
\begin{equation}
    P_{\alpha}(d_{k}|\vec{R})=\frac{P(d_k|H_{\alpha})R_{\alpha}}{\sum_{\beta}P(d_k|H_{\beta})R_{\beta}}\label{pastro}
\end{equation}
where, $H_{\alpha}$ is the hypothesis that a particular trigger has originated from the source category $\alpha$, $R_{\alpha}$ the expected  rate of \textit{detectable} triggers per unit time from source $\alpha\in\{\ac{BBH},\ac{BNS},\ac{NSBH},\text{Terrestrial}\}$, and $d_k$ the data associated with the $k^{th}$ trigger\footnote{Note that we express everything in terms of rates while ref.~\cite{multifgmc} used counts which relate to our rates as: $\Lambda_a=R_a\times T_{obs}$}. Note that the analysis of only \textit{detectable} triggers implies the existence of a threshold value for some detection statistic such as \ac{FAR} that is used to rank events by the GW search pipelines. 

In an offline analysis, wherein the data  corresponding to the full list of $N$ triggers found during an observing run is available, the probabilities in Eq.~\eqref{pastro} can be inferred self-consistently with the rates from the same dataset~\cite{fgmc,multifgmc}. This is achieved by first constructing the posterior distribution of the rates using the trigger data from a particular observing run~(say $r$), which looks like:
\begin{equation}
p(\vec{R}_r|\vec{d}_{N_r})\propto p(\vec{R}_r)e^{-\sum_{\alpha}R_{\alpha,r}T_{obs,r}}\prod_i^{N_r}\sum_{\alpha}P(d_i|H_{\alpha})R_{\alpha,r}T_{obs,r}\label{poterior}
\end{equation}
, where $\vec{d}_{N_r}$ is a set comprising data from all observed triggers, $N_r$ the total number of observed triggers, $T_{obs,r}$ the total observation time and $p(\vec{R}_r)$ an uninformative prior on the rates. Note that we are using the capitalized $P$ for probability and the regular $p$ for probability distribution/density, a notational choice made consistently throughout the rest of the paper. Once the rate posterior is constructed, it is possible to marginalize the probabilities in Eq.~\eqref{pastro} over the uncertainty in rate estimation:
\begin{equation}
    P_{\alpha}(d_k|\vec{d}_{N_r})=\int d\vec{R} P_{\alpha}(d_k|\vec{R}_r)p(\vec{R}_r|\vec{d}_{N_r})\label{pastro2}
\end{equation}
. Upon carrying out said marginalization, the probability of a GW trigger originating from a particular astrophysical source category $a\in\{\ac{BBH},\ac{BNS},\ac{NSBH}\}$ can be expressed in terms of the expectation values of the rates w.r.t. the posterior in Eq.~\eqref{poterior}, as in:
\begin{equation}
P_a(d_k|\vec{d}_{N_r})=\frac{\frac{P(d|H_{a})}{P(d|H_0)}\frac{\left<R_{a}\right>_{N_r/k}}{\left<R_0\right>_{N_r/k}}}{1+\sum_{b}\frac{P(d|H_{b})}{P(d|H_0)}\frac{\left<R_b\right>_{N_r/k}}{\left<R_0\right>_{N_r/k}}}\label{pastro3}    
\end{equation}
where, $\left<R_{\alpha}\right>_{N/k}=\int d\vec{R}p(\vec{R}|\vec{d}_{N/k})R_{\alpha}$ is the mentioned expectation value, $\vec{d}_{N/k}$  the dataset corresponding to all $N$ triggers except the $k^{th}$ one,  and the subscript $0$ representative of the noise/terrestrial source category. Note that the sum over $b$ in the denominator is over astrophysical categories only. Throughout the rest of this paper, Latin subscripts to rates and probabilities (such as $a,b,c$)  are chosen to represent an astrophysical source category while a Greek subscript (such $\alpha,\beta$) can denote astrophysical as well as terrestrial categories. In terms of the quantities defined in Eq.~\ref{pastro3}, the combined probability  of astrophysical origin can be written as $\PASTRO{}=\sum_a P_a$. For a derivation of Eq.~\eqref{pastro3} from Eq.~\eqref{pastro2} see ref.~\cite{multifgmc}. 

This framework has been used to compute offline \PASTRO{} values that were used in the generation of GW transient catalogs for the first three observing runs of LVK~\cite{gwtc-1,gwtc-2,gwtc-2.1,gwtc-3}. In the next subsection, we describe our rate approximation scheme that is needed to implement this framework accurately in low latency amidst an ongoing observing run.

\subsection{Computing $P_{a}$ in low latency: rate approximation}
\label{methods:rates}
As mentioned before, in order to evaluate Eq.~\eqref{pastro3}, the rate posterior needs to be constructed from the full list of triggers found during an observing run. However, for the purpose of assisting multi-messenger follow-up observations in real-time, $P_{a}$ has to be computed amidst an on-ongoing observing run i.e. in the absence of the complete list of triggers.

In such a scenario, our only option is to approximate the expected rates of detectable GW sources from their corresponding estimates yielded by past observing runs, while accounting for the change in sensitivity of the detectors in between the previous and ongoing runs. In particular, we only need approximate the ratio $\frac{\left<R_{a,r}\right>}{\left<R_{0,r}\right>}$ of the foreground and background rates since it is the only term involving them that appears in Eq.~\eqref{pastro}. To construct such an approximation scheme, we can exploit the following facts:
\begin{itemize}
    \item The astrophysical rate $\mathcal{R}$ of \acp{BBH}, \acp{BNS}, and \acp{NSBH}, per unit comoving volume per unit time, do not change in between observing runs.
    \item The rate of terrestrial triggers above some FAR threshold $(FAR_0)$ can be expected to be equal to said threshold: $R_{0}=FAR_{0}$.
\end{itemize}

Hence, estimates of $\mathcal{R}_{a}$ for $a\in\{\ac{BBH},\ac{BNS},\ac{NSBH}\}$ from previous observing runs can be used to approximate $R_{a,r}$ for an ongoing observing run $r$ given knowledge of the hypervolume~\cite{VT1} to which the detector is sensitive during said run:
\begin{equation}
    \left<R_{a,r}\right>=\left<\mathcal{R}_{a}\right>\frac{\left<VT\right>_{a,r}}{T_{\rm obs,r}}\label{rates}
\end{equation}
where, $\left< VT\right>_{a,r}$ is the sensitive hypervolume during observing run $r$ corresponding to the astrophysical source category $a$, $T_{obs,r}$ the corresponding live observation time, and $\left<\mathcal{R}_{a}\right>$ an estimate of the astrophysical rate yielded through the combined analysis of data from previous observing runs. However, if the ongoing observing run has a different time-volume sensitivity than the previous ones, $\left<VT\right>_{a,r}$ needs to be scaled appropriately from its values during past observing runs to account for said sensitivity change. While the sensitive hypervolume depends on various factors such as the source population model~\cite{VT1} (which itself remains uncertain to some degree~\cite{gwtc-3pop}), as a first-order step, simplistic scaling factors can be constructed by using the BNS horizon distance of the detector at various observing runs (including the ongoing one for which we use projections based on simulations~\cite{observingscenarios}). For example, if $d_{BNS,s}$ is the horizon distance of the detector network during observing run $s$ at an SNR threshold of 8 and $T_{obs,s}$ the live observation time, then $\left<VT\right>_{a,r}$ can be approximated in the following way:
\begin{equation}
    \left<VT\right>_{a,r}=\left<VT\right>_{a,old}\times \frac{d_{\rm BNS,r}^3T_{\rm obs,r}}{\sum_{s\in old}d_{\rm BNS,s}^3T_{\rm obs,s}}\label{V}
\end{equation}
where $old$ represents the combination of all past observing runs. The live observation times of the past runs $(T_{obs,s})$ can be approximated from the total number of terrestrial triggers found above threshold $(N_{0,old})$, their corresponding rate $(R_{0,old}=FAR_{0,old})$, and the individual run times $(T_{run,s})$ in the following way:
\begin{equation}
    T_{obs,s} = \frac{T_{\rm run,s}}{\sum_{s\in old} T_{\rm run,s}}\times \frac{N_{0,old}}{R_{0,old}}\label{T}
\end{equation}
Putting Eqs.~(\ref{rates},\ref{V},\ref{T}) together, we are able to construct an approximation for $\frac{\left<R_{a,r}\right>}{\left<R_{0,r}\right>}$ from estimates of various quantities available at the beginning of the ongoing observing run $r$, that takes the following form:
\begin{equation}
     \frac{\left<R_{a,r}\right>}{\left<R_{0,r}\right>}=\large\left<\mathcal{R}_a\right>\times \left<VT\right>_{a,old}\times \underbrace{\left(\frac{d_{\rm BNS,r}^3\sum_{s\in old}T_{\rm run,s}}{\sum_{s\in old}d_{\rm BNS,s}^3T_{\rm run,s}}\times \frac{FAR_{0,old}}{N_{0,old}FAR_{0,r}}\right)}_\textrm{\large$S_{r/old}$}\label{scaling}
\end{equation}
Notice here that $T_{obs,r}$ conveniently cancels in our self-consistent approximation scheme. The scaling factor on the RHS of Eq.~\eqref{scaling} is an improvement over the online \PASTRO{} calculations implemented upto O3 which assumed it to be 1 and hence did not account for the sensitivity change between O2 and O3, leading to potential underestimation of \PASTRO{} values. We demonstrate this through the analysis of O3 replay data in sec.~\ref{sec:results:rates} . We note that the rate approximation scheme described here is general in the sense that it can be used in conjunction with any low-latency matched filtering pipeline.
\subsection{Computing $P_{a}$ in low latency: category specific likelihood ratio}
\label{methods:LR}
With the self-consistent rate approximation scheme at our disposal, we can proceed to assemble a methodology for computing the remaining ingredients required for evaluating Eq.~\eqref{pastro3}, namely the marginalized likelihood ratios~( also known as Bayes factors in standard literature): 

\begin{equation}
K_{a}(d)=\frac{P(d|H_{a})}{P(d|H_{0})}    
\end{equation}
The Bayes factors quantify how much the data supports $H_{a}$ over $H_0$. In order to compute these quantities, one must model the population of CBCs in each astrophysical category using some fiducial function characterizing the distribution of measurable CBC parameters. Measurements of said parameters from detector strain data obtained by means of matched filtering can then be used to classify a potential GW candidate into various astrophysical categories given the different population models corresponding to each category.

Matched filtering search pipelines usually discretize the space of signal parameters to reduce computational cost, which results in a bank of template waveforms~\cite{findchirp}. A CBC signal is expected to \textit{ring up} templates in the bank with some SNR ~($\rho$). Multi-detector coincidences are formed when identical templates are triggered at different observatories within the propagation time of GWs between said observatories. The triggered templates are usually ranked in significance with some statistic (say $x$) that is often computed from additional quantities assigned to GW triggers by the search pipeline.

Among the triggers that pass a chosen detection threshold $x_{th}$ in $x$, either the highest $x$ or highest $\rho$ template within a certain time-window is chosen to represent a potential GW candidate that might have occurred in said window. The data corresponding to a particular GW candidate $(d_k)$ relevant for Bayes factor calculation thus comprises three quantities assigned to it by the search pipeline namely the representative template ($t_k$), the corresponding  ranking statistic ($x_k$) and SNR $\rho_k$. 

To compute the Bayes factor accurately from the trigger data, one needs to account for the possibility of random noise fluctuations causing a true signal originating from source category $a$ to be recovered in a template whose parameters correspond to a different source category $b$. This is particularly true if the different source categories map onto neighboring regions of the template bank separated by some hard boundary.  

Hence, a true signal corresponding to template $t_j$ and SNR $\rho$, can have a finite probability ($P(t_k|t_j,\rho)$) of being found in a different template $t_k$, depending on how similar said templates are with each other. In terms of this probability and a category-specific distribution of template parameters, it is possible to evaluate the likelihood with which any template $t_k$ in the bank might be triggered by a CBC signal originating from source category $a$:

\begin{equation}
    P(t_k,\rho|H_a)= \sum_j  P(t_k|t_j,\rho)p(\vec{\theta}(t_j)|H_a)p(\rho|H_a) V_j  \label{mass-model}
\end{equation}
where, $\vec{\theta}(t_j)$ is the CBC parameters corresponding to the template $t_j$, $V_j$ the volume in parameter space associated with $\vec{\theta}(t_j)$  and $p(\rho|H_a)$ the prior distribution of SNRs corresponding to sources in category $a$~\cite{Fong:2018}. 

To compute $P(t_k|t_j,\rho)$, we make the assumption that migration between true and recovered templates is solely due to Gaussian fluctuations~\cite{Fong:2018}. Furthermore, since the purpose of these \textit{template-weights} ($P(t_k,\rho_k|H_a)$) is to account for the miss-classification of GW sources due to the template migration caused by noise, it is safe to assume, for a first approximation, that $\rho_k = \rho_j$. Under these considerations, the probability of template migration takes the following form:

\begin{equation}
    P(t_k|t_j,\rho)=\frac{1}{\sqrt{2\pi}}e^{-\frac{1}{2}\rho^2\left[1-2t_k\cdot t_j+(t_j\cdot t_k)^2\right]}\label{eq:confusion}
\end{equation}
where $t_k\cdot t_j$ is the noise averaged inner product of the normalized template waveforms. For a derivation of Eq.~\eqref{eq:confusion} see \ref{appendix:tw1}.

The volume element $V_j$ corresponding to each template needed for evaluating Eq.~\eqref{mass-model} can often be pre-computed from the template bank. For an arbitrary bank, it is possible to approximate the volume element $V_j$  corresponding to a template $t_j$ as the volume of the Voronoi cell~\cite{Voronoi1,Voronoi2} that contains the point $\vec{\theta}(t_j)$~\cite{Fong:2018}. However the method of Voronoi decomposition is computationally expensive and it is preferable to compute $V_j$ as part of bank generation itself, if possible, as in the case of banks populated using methods like~\cite{hanna2022binary}. Since different search pipelines use different template banks and ranking statistics, the remainder of the Bayes factor calculation will also vary from pipeline to pipeline. We demonstrate how such a calculation can be implemented within the \GSTLAL{} search pipeline~\cite{messick2016gstlal,cannon2020gstlal,sachdev2019gstlal} while noting that our formalism can also be applied to other pipelines by simply replacing the \GSTLAL{}-specific part of the Bayes factor calculation that is described below.

The O4 template bank used by the \GSTLAL{} search pipeline places templates in the $\log_{10}\left(m_1\right) \times \log_{10}\left(m_2\right) \times \chi_\text{eff}$ space by splitting up the parameter space via a binary tree approach such that the volumes of the nearby templates are similar. Here, $\chi_\text{eff}$ is $\frac{m_1 \times s_{1,z} + m_2 \times s_{2,z}}{m_1 + m_2}$ where $m_i$ and $s_{i,z}$ are the masses and z-component spins of the $i^{\text{th}}$ component of the binary. The volumes surrounding templates are computed using the minimal match of the templates allowed in the template bank. For details regarding the volume calculation for the template bank used in O4 by the \GSTLAL{} search pipeline see Ref.~\cite{bank}. 

Putting all of these pieces together yields a robust estimate of $P(t_k,\rho_k|H_a)$, one that accounts for the mismatch between true and recovered waveforms due to detector noise. The variation of these quantities as a function of SNR and template parameters are summarized in \ref{appendix:tw2}. We find said variation to be consistent with our expectations regarding miss-classification among various categories at different SNRs. 

We note that \GSTLAL{}'s online analyses upto O3 did account for this migration between true and recovered templates accurately and classified sources into astrophysical categories either based solely on the recovered template parameters or based on simulation campaigns that were not only expensive to generate but also sparse compared to the full template bank. We demonstrate the accuracy of the source classification resulting from the use of our template weights in sec~\ref{sec:results-weights}.

With the template weights accounting for miss-classification among astrophysical categories, the  Bayes factors can be computed from said template weights and the probability distributions of the search pipeline's ranking statistic conditional on the signal and noise hypotheses ($H_1$ and $H_0$)~\cite{multifgmc}.  For the \GSTLAL{} search pipeline, its ranking statistic is itself the log of the likelihood ratio: $e^{x_k}\propto P(d_k|H_1)/P(d_k|H_0)$~\cite{cannon2015likelihoodratio,messick2016gstlal,hanna2020likelihoodratio,likelihood_ratio}. Hence, in terms of \GSTLAL{}'s ranking statistic ($x_k$), the Bayes factors can be written in the following way:

\begin{eqnarray}
     K(x_k,\rho_k,t_k)=A(FAR_0)e^{x_k}\times\frac{P(t_k,\rho_k|H_{a})}{\left(\frac{1}{3}\right)\sum_{b}P(t_k,\rho_k|H_{b})}\label{bayesfac2}\\
    A(FAR_{0})=\left[\int_{x(FAR_0)} ^{\infty}dxe^xp(x|H_0)\right]^{-1}
\end{eqnarray}
where $x(FAR_0)$ is the log-likelihood ratio corresponding to the FAR threshold used throughout the \PASTRO{} calculation and $p(x|H_0)$ is the normalized probability distribution of the ranking statistic conditional on the noise hypothesis. For a derivation of Eq.~\eqref{bayesfac2}, see ~\ref{appendix:lr}.

With the Bayes factors calculable on the fly from trigger data and the rates pre-computed using the scaling relation~\eqref{scaling}, robust and accurate \PASTRO{} values can be evaluated in low-latency through Eq.~\eqref{pastro3} for assisting in multi-messenger follow-up observations to GW triggers in real-time. The code developed to implement this analysis is publicly available in the python package \textsc{gwsci-pastro}\footnote{https://pypi.org/project/gwsci-pastro/}
\section{Illustrative results: mock data challenge}
\label{sec:results}
We have tested our formalism for calculating \PASTRO{} through \GSTLAL{}'s participation in a rigorous \ac{MDC}~\cite{mdc_analytics,ewing2023performance}. The MDC involved a replay of strain data from the detectors at Hanford, Livingston and Virgo spanning 40 days of O3, from \MDCSTART{} to \MDCEND{}, along with simulated signals injected into the data stream. Several low-latency search pipelines (including \GSTLAL{}) were used to analyze this mock dataset with our new \PASTRO{} calculation implemented as part of \GSTLAL{}'s analysis. For further details regarding the MDC see refs~\cite{mdc_analytics,ewing2023performance}. 

To validate our formalism for calculating \PASTRO{}, we first demonstrate the improvements resulting from our rate-approximation scheme in the context of real GW events replayed in the MDC. Afterward, we use the \PASTRO{} values of simulated events computed using our formalism to provide a rigorous validation for the developed rate approximation and source classification schemes.
\subsection{GW events replayed in the MDC}
\label{sec:results:rates}
As mentioned before, the real-time \PASTRO{} calculations necessitate the use of rate approximation schemes to compensate for the absence of the full list of triggers found during an observing run, which can only be available in post-processing. In that sense, our online \PASTRO{} values are approximations to the ones computed self-consistently with the rates in an offline analysis carried out at the end of the observing runs for generating GW transient catalogs. Therefore, as a validity test of our rate-approximation scheme, \PASTRO{} values computed using our formalism for O3 GW candidates replayed in a MDC can be compared with results from the O3 offline analysis.

For this MDC, we can thus pretend that O3 is the ongoing observing run while O1 and O2 are the previous ones. To compute rate estimates for calculating \PASTRO{} values in the MDC, we therefore scale the sensitive hyper-volume of O1 and O2  obtained from offline analyses, using the projections of the O3 BNS ranges available at the beginning of O3~\cite{observingscenarios}. We then use the estimates of the astrophysical rates available from offline analyses of O1 and O2 in conjunction with the scaled hypervolumes to find $\left<R_{a,O3}\right>/\left<R_{0,O3}\right>$. 

The accuracy of our rate approximation can be demonstrated by comparing the aforementioned scaled rate estimates to the unapproximated ones obtained from an offline analysis of public LVK data corresponding to O3 triggers found by \GSTLAL{}~\cite{O3-data-release}. The comparison is summarized in table~\ref{table:rates-mdc}. It can be seen that the approximated scaling relation yields BBH rates that are remarkably accurate given how close they are to the offline results. On the other hand, the unscaled O1O2 rates are smaller by a factor of 2. 

The scaled estimates of BNS and NSBH are further away from the offline O3 estimates as compared to BBH, which is to be expected due to the following reasons. The number of confident BNS and NSBH detections in O1O2 are 1 and 0 respectively~\cite{gwtc-1}. Similarly, the corresponding numbers in O3 are 1 and 2~\cite{gwtc-2, gwtc-3}. Hence, both sets of rate estimates, namely the ones scaled from O1O2 and the ones yielded by the O3 offline analysis, are subject to much larger Poisson uncertainties for BNS and NSBH than for BBH. The BNS and NSBH rate estimates for O4~(listed in table~\ref{table:rates-o4}), which are scaled from the O1O2O3 rates and VTs, can be expected to be more accurate given the larger number of BNS and NSBH detections in O1O2O3 as compared to O1O2.

\begin{table}[h]
\centering
\begin{tabular}{cccc}
\hline
\hline
Category & Unscaled & Scaled & Offline \\
\hline
BBH & $3.03\times 10^{-3}$ & $7.67\times10^{-3}$ & $7.80\times10^{-3}$\\
\hline
BNS & $4.23 \times 10^{-4}$ &$1.09\times10^{-3}$ &  $1.7\times 10^{-4}$\\
\hline
NSBH & $3.04\times 10^{-4}$ & $7.87\times 10^{-4}$ & $5.9\times 10^{-4}$\\
\hline
\end{tabular}
\centering
\caption{
Comparison of $\left<R_{a,O3}\right>/\left<R_{0,O3}\right>$ for the three cases (unscaled, scaled and O3 offline) described above. }
\label{table:rates-mdc}
\end{table}

Furthermore, table~\ref{table:rates-mdc} is indicative of the fact that using the unscaled rates and hypervolumes of O1 and O2 has led to the underestimation of \PASTRO{} values in the O3 online analysis. To verify this claim, we compare the \PASTRO{} of real GW events replayed in the MDC computed using the 3 different rate estimates listed in table~\ref{table:rates-mdc}. We choose the fiducial population model required for computing the template-weights (and hence \PASTRO{}) to be a Salpeter function~\cite{salpeter1955} in primary mass and uniform in all other parameters:
\begin{equation}
    p(m_1,m_2,\chi_{1z},\chi_{2z}|H_a)\propto m_1^{-2.35}\frac{\Theta(m_1-m_2)}{m_1-m_{min}}\begin{cases} 1 & (m_1,m_2,\chi_{1z},\chi_{2z})\in S_a\\
    0 & o.w.       
    \end{cases}\label{astromass-model}
\end{equation}
. Here $\Theta$ is the Heaviside step function, $m_{min}$ is the minimum mass of templates in the bank and $S_a$ are hyper-cubes in the space of the parameters $(m_1,m_2,\chi_{1z},\chi_{2z})$ whose edges correspond to category specific upper and lower boundaries. The boundaries in parameter space separating BNS from NSBH and NSBH from BBH are determined by the maximum NS mass and spin which are taken to be $3~M_{\odot}$ and $0.5$ respectively. Masses and spins of BHs are allowed to go up to $300~M_{\odot}$ and $0.99$ respectively.

\begin{figure*}[htt]
\centering
\includegraphics[width=0.8\textwidth]{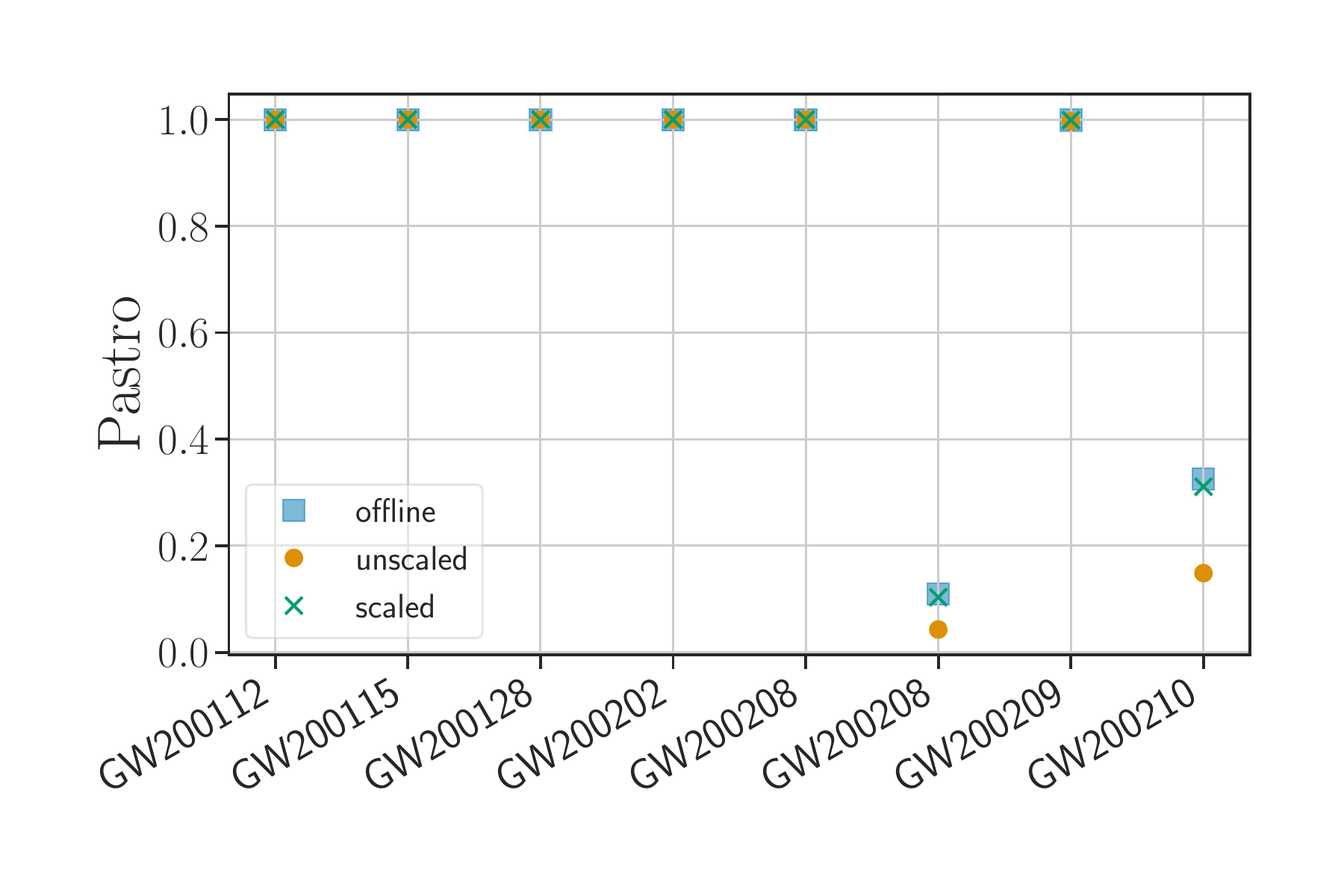}
\caption{\label{fig:pastro-dates} Comparison of \PASTRO{} values computed using scaled, unscaled and offline rate estimates. The x-axis marks the actual dates on which the replayed GW events were discovered in O3.}
\end{figure*}

With $P(t_k,\rho_k|H_a)$ pre-computed using Eq.~\eqref{mass-model} for the population model in Eq.~\eqref{astromass-model}, \PASTRO{} values for the real GW events replayed in the MDC can be computed for validating our rate-approximation schemes. For a visual representation of how the \PASTRO{} values computed using the aforementioned template weights and the 3 different rates listed on table~\ref{table:rates-mdc} compare against each other see Fig.~\ref{fig:pastro-dates}. 

It can be seen that while both the scaled and unscaled rate estimates yield accurate \PASTRO{} values for the highly significant candidates, the scaled rate estimates perform significantly better for the marginal events than the unscaled ones. In particular, it can be seen that the low-latency \PASTRO{} of the NSBH candidate GW200210\_092254 would be underestimated by a factor of 2 if not for our accurate scaling approximation. 

Given this performance in the MDC, we conclude that our formalism can be expected to yield \PASTRO{} values of real GW events in \GSTLAL{}'s online analysis of O4 data, that approximate the results of a corresponding offline analysis with higher accuracy, as compared to what was implemented in O3. In sec.~\ref{sec:results-weights}, We provide a more rigorous demonstration of the necessity of our rate scaling approximations through analysis of the set of simulated sources in the MDC.

The approximate rates to be used by \GSTLAL{} to calculate online \PASTRO{} values during O4 are listed in table~\ref{table:rates-mdc}. They are largely consistent with the estimates yielded by Monte Carlo simulations carried out for assessing LVK's observing capabilities in O4~\cite{observingscenarios}. The rates in table~\ref{table:rates-o4} shall be used in conjunction with the template-weights corresponding to the Salpeter population-model in Eq.~
\eqref{astromass-model} by \GSTLAL{} in O4 for computing robust and accurate \PASTRO{} values in its online analysis of O4.

\begin{table}[h]
\centering
\begin{tabular}{cccc}
\hline
\hline
Category & $\left<\mathcal{R}_{a}\right>$ & $\left<VT_{a,old}\right>$ & $\left<R_{a,O4}\right>/\left<R_{0,O4}\right>$  \\
\hline
BBH & $40~Gpc^{-3}yr^{-1}$ & $1.6~Gpc^3yr$ & 0.018\\
\hline
BNS & $163~ Gpc^{-3}yr^{-1}$ & $0.014~Gpc^3yr$ & $6.7\times 10^{-4}$\\
\hline
NSBH & $68.0~ Gpc^{-3}yr^{-1}$ & $0.048~ Gpc^3yr$ & $9.1\times 10^{-4}$\\
\hline
\end{tabular}
\centering
\caption{
The estimates of $\left<R_{a,O4}\right>/\left<R_{0,O4}\right>$ to be used by \GSTLAL{} in the O4 online analysis. The columns $\left<\mathcal{R}_{a}\right>$ and $\left<VT_{a,old}\right>$ list estimates of the astrophysical rates and the sensitive hyper-volume obtained from the offline analysis of O1O2O3 (appendix C3 of ref.~\cite{gwtc-3pop}).}
\label{table:rates-o4}
\end{table}

We note that the distribution in Eq.~\eqref{astromass-model} is a straw-person representative of the true CBC mass and spin population which remains uncertain to this date~\cite{gwtc-3pop}. In particular, the population-level distributions of masses and spins of both \ac{BNS} and \ac{NSBH} systems are subject to significant Poisson uncertainties given the small number of detections made to date. If new \acp{BNS} and \acp{NSBH} discovered in the first half of \ac{O4} lead to more precise measurements of their population properties and if the new measurements are found to deviate significantly from our straw-person population model, the template weights corresponding to some best estimate of the newly measured population can be re-computed straightforwardly leading to even more accurate source classification in the second half of \ac{O4}.

However, improvements in our knowledge of the true CBC population can lead to better classification if and only if the deviation between true and recovered signal parameters due to noise fluctuations is correctly accounted for within the formalism of calculating \PASTRO{}. We thus want to test for the accuracy with which our formalism can correctly account for the probability of miss-classification among astrophysical source categories given knowledge of true CBC population distribution.

To perform such a validation study, we calculate the category-specific \PASTRO{} values~($P_a$) for the simulated events in the \ac{MDC} that comprise a known fiducial population of CBCs. We summarize the results obtained from said study in the next subsection. 

\subsection{Simulated sources in the MDC}
\label{sec:results-weights}

The category-specific \PASTRO{} values of the simulated sources that are  injected into the replay data as part of the MDC, can be used to validate our source classification and rate-approximation schemes. The template weights as well as the rate of injected events per comoving volume, both of which are needed for calculating $P_a$, will depend on the population-level distributions of CBC parameters that were used to generate the mentioned injections.

The masses and spins characterizing the simulated signals in the \ac{MDC} were drawn from fiducial population-level distributions~(one corresponding to each source category), the functional forms of which were chosen to be truncated power-laws in masses and uniform in spins~\cite{mdc_analytics}, as in:
\begin{equation}
    p(m_1,m_2,\chi_{1z},\chi_{2z}|H_a)\propto m_1^{-\beta_{1,a}}m_2^{\beta_{2,a}}\Theta(m_1-m_2)\begin{cases} 1 & (m_1,m_2,\chi_{1z},\chi_{2z})\in S_a\\
    0 & o.w.       
    \end{cases}\label{injmass-model}
\end{equation}
. Here, $S_a$ are hyper-cubes in the space of the parameters $(m_1,m_2,\chi_{1z},\chi_{2z})$ whose edges correspond to category-specific upper and lower boundaries. The redshift distribution of these injections is taken to be uniform in co-moving volume upto a category-specific maximum ($z_{max,a}$)~\cite{mdc_analytics}. The values of the power-law indices and the maximum redshifts corresponding to each category are listed in table~\ref{table:inj-hyperparams}. As before, the boundaries in parameter space separating BNS from NSBH and NSBH from BBH are determined by the maximum NS mass and spin. For the population of simulated sources used in the MDC, these values were taken to be $2.05~M_{\odot}$ and $0.4$ respectively. Masses and spins of BH components were allowed to go upto $100~M_{\odot}$ and $0.99$ respectively for BBHs and upto $60~M_{\odot}$ and $0.99$  for NSBHs while no NSs lighter than $1{M_{\odot}}$ were drawn.

\begin{table}[h]
\centering
\begin{tabular}{cccc}
\hline
\hline
Category & $\beta_{1}$ & $\beta_2$ &$z_{max}$ \\
\hline
BBH & $2.35$ & $1$ & $1.9$\\
\hline
BNS & $0$ & $0$ & $0.15$\\
\hline
NSBH & $-1$ & $0$ & $0.25$\\
\hline
\end{tabular}
\centering
\caption{The powerlaw indices characterizing the mass-distribution of simulated sources corresponding to each category.}
\label{table:inj-hyperparams}
\end{table}

Given these population distributions, the template weights corresponding to each of them can be computed using Eqs.~\eqref{mass-model}. The variation of these template weights with SNR, masses, and spins are discussed in \ref{appendix:tw2}. 

To compute \PASTRO{} for the simulated sources, we also need the rate of injections which is significantly higher than the astrophysical rates listed in table~\ref{table:rates-mdc}. The rate of detectable injections can be estimated from the total number of injections drawn and the maximum redshift corresponding to each category, as in: 

\begin{equation}
    \frac{\left<R_{r,a}\right>}{\left<R_{0}\right>}|_{inj}=\frac{N_{inj,a}}{V(z_{max,a})T_{inj}}\left<VT_{old,a}\right>S_{r/old}\label{inj-rate}
\end{equation}
, where $N_{inj,a}$ is the total number of injections drawn from category $a$, $S_{r/old}$ the scale factor derived in Eq.~\eqref{rates} and $V(z)$ is the co-moving volume upto redshift $z$. Armed with the rates in Eq.~\eqref{inj-rate} and the template weights corresponding to the population distributions in Eq.~\eqref{injmass-model}, we can compute \PASTRO{} for all the simulated sources in the MDC that are found by the search pipeline. We can then compare the cumulative sum of \PASTRO{} values with the number of found injections for validating our rate-approximation and source classification schemes.

\begin{figure}[htp]
    \includegraphics[width=0.9\textwidth]{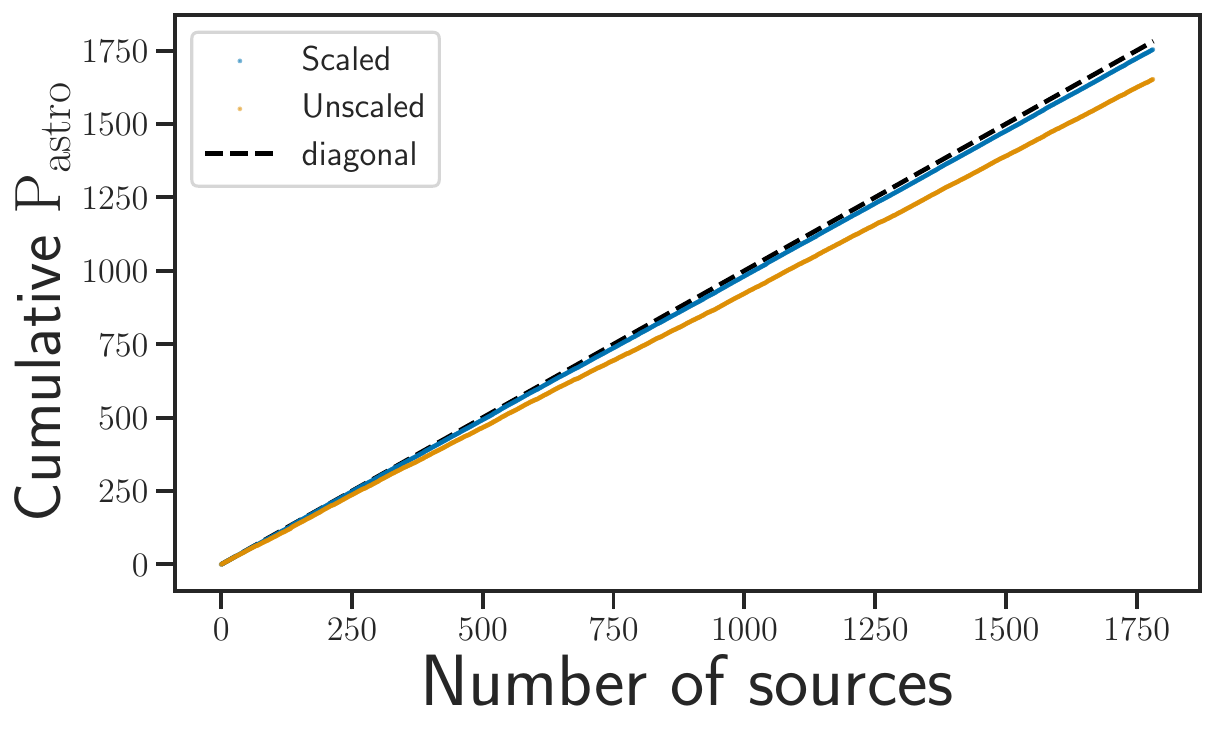}
\caption{\label{fig:pastro-injections-pp} Recovery of simulated sources. The y-axis represents the cumulative sum of \PASTRO{} values of all the triggers found above threshold by the search pipeline. The blue line corresponds to the scaled rate estimates in Eq.~\eqref{inj-rate} while the orange curve corresponds to un-scaled rate estimates with $S_{new/old}=1$}
\end{figure}

To demonstrate the improvements resulting from our rate-approximation scheme, we compute two sets of \PASTRO{} values, one using rate estimates obtained from the correct values of $S_{r/old}$ and the other using rates that correspond to $S_{r/old}=1$. We then compare the cumulative sum of \PASTRO{} values $(\sum_i\sum_a P_a(d_i))$ corresponding to both sets of rate estimates against the total number of found injections. The results of said comparison are summarized in Fig.~\ref{fig:pastro-injections-pp}.

In the case of perfect signal recovery, the plotted curve would lie exactly atop the diagonal. Fig.~\ref{fig:pastro-injections-pp} thus validates that our formalism yields \PASTRO{} values that correspond to a near-perfect recovery of all simulated sources with very few injections lost to terrestrial. It also demonstrates how the rate-approximation part of our formalism prevents the underestimation of \PASTRO{} since the \PASTRO{} values generated using un-scaled rate estimates are significantly below the diagonal than the scaled ones.

\begin{figure}[H]
\centering
\includegraphics[width=0.9\textwidth]{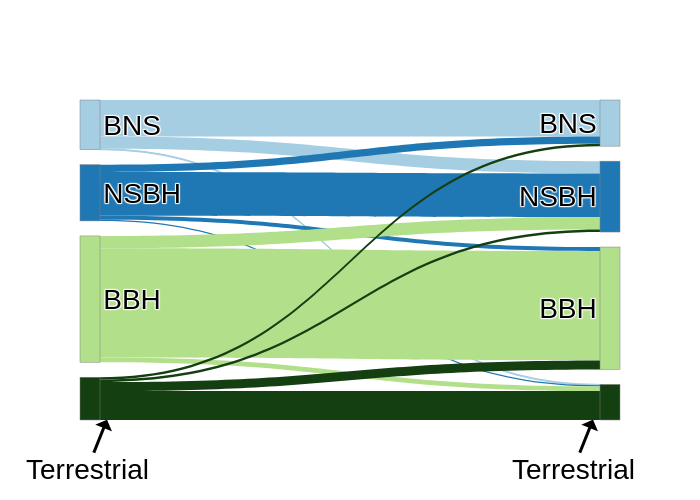}
\caption{\label{fig:pastro-injections-sankey} Classification of simulated sources. In this Sankey diagram, the size of each link is representative of $f(b|a)$ with $a$ on the left and $b$ on the right. }
\end{figure}

To now demonstrate the accuracy with which our template weights can classify triggers into various astrophysical categories we compute the fraction of simulated sources from each category $a$ that are classified into category $b$. The value of this fraction can be obtained from the cumulative sum of category-specific \PASTRO{} values of the simulated events in the following way:

\begin{equation}
    f(b|a)=\frac{\sum_{i\in a}P_b(d_i)}{N_a}
\end{equation}
where $i\in a$ represents the the trigger corresponding to the $i$th simulated event from source category $a$ and $N_a$ is the total number of simulated events from $a$ that are recovered by the search pipeline with $FAR\geq FAR_0$. To compute the fraction of simulated sources lost to the terrestrial category, we use $f(\mathrm{Terrestrial}|a)=\left[\sum_i(1-\sum_b P_b(d_i))\right]/N_a$.For the simulated astrophysical sources and the noise triggers in the MDC, this set of fractions obtained from our \PASTRO{} estimates are represented in Fig.~\ref{fig:pastro-injections-sankey}.

We can see in Fig.~\ref{fig:pastro-injections-sankey} that the majority of simulated sources are correctly classified with a very small fraction of miss-classifications. For the \ac{BNS} injections, we find that $P_{\rm BNS}$, $P_{\rm NSBH}$ and $P_{\rm Terrestrial}$ accounts for 73\%, 25\% and 2\% of them respectively with a negligible fraction lost to \ac{BBH}. Similarly, for the \ac{NSBH} injections, we find that 77\%, 13\% and 8\% are classified as \ac{NSBH}, \ac{BNS}, and \ac{BBH} respectively with only 2\% lost to terrestrial. 86\% \acp{BBH} are correctly classified with 10\% and 4\% injections recovered as \ac{NSBH} and terrestrial. Note that very few of the potentially EM bright sources (BNS and NSBH) are lost to BBH or terrestrial.

To summarize, we conclude that Figs~\ref{fig:pastro-injections-sankey} and \ref{fig:pastro-injections-pp} are implicative of the following. In a simulated universe with known populations of BBH BNS and NSBH systems that merge at some known astrophysical rate per comoving volume per time, our formalism can be used to compute \PASTRO{} in real-time that enable highly accurate source classification in the presence of noisy data. Furthermore, we conclude that the rate approximation scheme developed as part of our formalism is a necessity since a significant fraction of simulated sources has been shown to be misclassified as terrestrial without it.  

\section{Conclusion and future prospects}
\label{sec:conclusion}
We have developed a self-consistent analysis framework capable of accurately classifying GW candidates in real-time, for the purpose of assisting multimessenger follow-up observations during O4. We have implemented our formalism through the \GSTLAL{} search pipeline while noting that our rate approximation scheme is applicable to other pipelines as well. We have demonstrated the accuracy of our classification scheme and the improvements resulting from our rate approximation scheme, both through the analysis of real and simulated GW data as part of an MDC. With the robust and accurate \PASTRO{} values yielded by our framework, the \GSTLAL{} search pipeline can be expected to better assist in multimessenger follow-up campaigns to the GW alerts sent out during O4, as compared to O3.

While our formalism is an improvement over existing frameworks for calculating \PASTRO{}, further developments leading to even better source classification can potentially be implemented in the near future. For example, even though our template weights facilitate the correct classification of most GW candidates, they are still based on point estimates of CBC parameters computed by the search pipelines and hence susceptible to potential biases. On the other hand, Bayesian parameter estimation~(PE) has been shown to yield estimates of these quantities that are significantly more accurate than the point estimates computed by search pipelines. Given the recent developments in the field of low-latency PE in real-time~\cite{Planckow2015,rose2022supplementing}, it is possible to replace our template-weights-based classification scheme with one that makes use of more accurate mass and spin measurements yielded by said PE. The related developments and the associated validation studies are part of an ongoing exploration.

To summarize, we have implemented several improvements in the low-latency computation of category-specific \PASTRO{} values by the \GSTLAL{} search pipeline for assisting in the search for counterparts to GW candidates in O4. Given the performance of our analysis framework in an MDC comprising real and simulated GW data, we conclude that these improvements will lead to the correct identification and classification of several GW candidates in O4 that would have otherwise been marked as terrestrial or miss-classified into an incorrect astrophysical category. While our classification scheme is susceptible to uncertainties in the population model, these uncertainties are expected to decrease as more and more events are found, leading to further improvements in classification in the later parts of O4 and beyond.
\section{Acknowledgement}
 The authors are grateful for computational resources provided by the LIGO
    Laboratory and supported by National Science Foundation Grants PHY-0757058
    and PHY-0823459.  This material is based upon work supported by NSF's LIGO
    Laboratory which is a major facility fully funded by the National Science
    Foundation.  LIGO was constructed by the California Institute of Technology
    and Massachusetts Institute of Technology with funding from the National
    Science Foundation (NSF) and operates under cooperative agreement
    PHY-1764464. The authors gratefully acknowledge the Italian Istituto Nazionale 
    di Fisica Nucleare (INFN), the French Centre National de la Recherche Scientifique 
    (CNRS) and the Netherlands Organization for Scientific Research, for the construction 
    and operation of the Virgo detector and the creation and support  of the EGO consortium.  
    The authors are grateful for computational resources provided
    by the Pennsylvania State University's Institute for Computational and Data
    Sciences (ICDS) and the University of Wisconsin Milwaukee Nemo and support
    by NSF PHY-\(2207728\), NSF PHY-\(2011865\), NSF OAC-\(2103662\), NSF PHY-\(1626190\), and NSF
    PHY-\(1700765\).  This paper
    carries LIGO Document Number LIGO-P2300141. \\

\appendix
\section{Derivation of Bayes Factors from \GSTLAL{}'s ranking statistics}
\label{appendix:lr}
In Sec.~\ref{methods:LR}, we mentioned that among the various quantities assigned to a GW candidate by a search pipeline, the trigger data relevant to Bayes factor calculation can be thought to comprise only three quantities, namely the ranking-statistic, the representative template, and the SNR. For the \GSTLAL{} search pipeline, Eq.~\eqref{bayesfac2} suggests that it is indeed the case. However, in order to derive Eq.~\eqref{bayesfac2} itself, the details of how the ranking statistic is computed from additional pipeline-assigned quantities will become relevant. 

For \GSTLAL{}, the full dataset corresponding to a particular trigger comprises additional quantities to $t_k$ and $\rho_k$ such as the signal-based veto parameter $\vec{\xi^2}$ at each detector. The other independent quantities that comprise the full dataset are measurements of arrival time and coalescence phase $(\vec{\Delta t}_k,\vec{\Delta \phi}_k)$ by different detectors expressed relative to the corresponding measurements by some reference detector ($t_{\rm ref,k},\phi_{\rm ref,k}$). The log-likelihood ratio used by \GSTLAL{} to rank triggers can be expressed in terms of the probability distributions of these quantities~\cite{hanna2020likelihoodratio,likelihood_ratio} in the following way:

\begin{equation}
    e^{x_k}\propto \frac{p(\vec{\xi}_k^2,\vec{\Delta t}_k,\vec{\Delta \phi}_k,t_{\rm ref,k},\phi_{\rm ref,k},\vec{O}_k|\rho_k,t_k,H_1)}{p(\vec{\xi}_k^2,\vec{\Delta t}_k,\vec{\Delta \phi}_k,t_{\rm ref,k},\phi_{\rm ref,k},\vec{O}_k|\rho_k,t_k,H_0)}\times \frac{P(t_k,\rho_k|H_1)}{P(t_k,\rho_k|H_0)}\label{lnlra1}
\end{equation}
where $\vec{O}$ is the subset of detectors that found the trigger with a \ac{SNR} above threshold. While we do describe in Sec.~\ref{methods:LR} the calculation of $P(t_k,\rho_k|H_1)$,  details regarding the other individual probabilities in Eq.~\eqref{lnlra1} can be found in ref.~\cite{hanna2020likelihoodratio}. On the other hand,  in terms of the same trigger-data used for calculating $x_k$, the Bayes factors required for \PASTRO{} can be written as:
\begin{equation}
    K_a(d_k)=\frac{p(\vec{\xi}_k^2,\vec{\Delta t}_k,\vec{\Delta \phi}_k,t_{\rm ref,k},\phi_{\rm ref,k},\vec{O}_k|\rho_k,t_k,H_a)}{p(\vec{\xi}_k^2,\vec{\Delta t}_k,\vec{\Delta \phi}_k,t_{\rm ref,k},\phi_{\rm ref,k},\vec{O}_k|\rho_k,t_k,H_0)}\times \frac{P(t_k,\rho_k|H_a)}{P(t_k,\rho_k|H_0)}\label{bfa1}
\end{equation}
. We note that template $t_k$ is the only parameter that determines which astrophysical source category the signal might have originated from in the sense that the only difference between astrophysical source categories is the population-level distributions of the template parameters corresponding to each category. Hence the probabilities of the remaining quantities $(\vec{\xi}_k^2,\vec{\Delta t}_k,\vec{\Delta \phi}_k,t_{\rm ref,k},\phi_{\rm ref,k},\vec{O}_k)$ conditional on $H_a$ can be expected to equal the probabilities of the same quantities conditioned instead on $H_1$. Under this consideration, we can use Eq.~\eqref{lnlra1} to rewrite Eq.~\eqref{bfa1} in the following way:
\begin{equation}
    K_a(d_k) = Ae^{x_k}\frac{P(t_k,\rho_k|H_a)}{P(\rho_k,t_k|H_1)}\label{bfa2}
\end{equation}
Since the category-specific signal hypotheses $(H_a)$ correspond to disjoint regions of the template bank ($H_a\cap H_b=\emptyset\quad for \quad a\neq b$) that together span the entire bank ($\cup_a H_a=H_1$), we can re-write the signal model in the denominator to be $P(\rho_k,t_k|H_1)\approx \sum_a P(t_k,\rho_k|H_a)P(H_a|H_1)$, where  $P(H_a|H_1)\approx \frac{1}{3}$ for $a\in\{\ac{BBH},\ac{BNS},\ac{NSBH}\}$. Under this consideration, Eq.~\eqref{bfa2} can be expressed as follows:

\begin{equation}
     K_a(d_k) = Ae^{x_k}\frac{P(t_k,\rho_k|H_a)}{(\frac{1}{3})\sum_b P(t_k,\rho_k|H_b)}\label{bfa3}
\end{equation}
To obtain an expression for the normalization constant A, we can make use of \GSTLAL{}'s estimates of the probability distributions of its ranking statistic. The probability distribution of obtaining some value for the ranking statistic $x$ can be obtained from the same probabilities used to calculate the ranking statistic itself conditional on either hypothesis $H_1~or~H_0$, as follows:

\begin{equation}
    p(x|H_{(0,1)})=\sum_{j}\int_{\Sigma(x)} d\Sigma^{N-1}p(\vec{\xi}_k^2,\vec{\Delta t}_k,\vec{\Delta \phi}_k,t_{\rm ref,k},\phi_{\rm ref,k},\vec{O}_k|\rho_k,t_k,H_{(0,1)})P(t_j,\rho_k|H_{(0,1)})\label{rankstatpdfa1}
\end{equation}
where $\Sigma(x)$ is a surface in the space of the parameters $(\rho_k,\vec{\xi}_k^2,\vec{\Delta t}_k,\vec{\Delta \phi}_k,t_{\rm ref,k},\phi_{\rm ref,k},\vec{O}_k)$ corresponding to a fixed value of the ranking statistic $x$, as defined by Eq.~\eqref{lnlra1}. For details regarding how the integrals are implemented by means of importance sampling, see ref.~\cite{cannon2015likelihoodratio,messick2016gstlal}. Using Eq.~\eqref{lnlra1}, we can re-write Eq.~\eqref{rankstatpdfa1} for the ranking statistic distribution conditional on the signal hypothesis in terms of that conditional on the noise hypothesis in the following way:

\begin{equation}
     p(x|H_{1})=Ae^{x}\sum_{j} \int_{\Sigma(x)} d\Sigma^{N-1}p(\vec{\xi}_k^2,\vec{\Delta t}_k,\vec{\Delta \phi}_k,t_{\rm ref,k},\phi_{\rm ref,k},\vec{O}_k|\rho_k,t_k,H_0)P(t_j,\rho|H_{0})\label{rankstatpdfa1.5} 
\end{equation}

\begin{equation}
    \implies p(x|H_{1}) = Ae^{x}p(x|H_0)\label{rankstatpdfa2} 
\end{equation}
Now demanding that both $p(x|H_1)$ and $p(x|H_0)$ is normalized, we can find an expression for A in terms of the noise pdf:
\begin{equation}
     A=\left[\int_{x_{min}} ^{\infty}dxe^xp(x|H_0)\right]^{-1}
\end{equation}
where $x_{min}$ is the ranking statistic corresponding to the FAR threshold above which \PASTRO{} values are calculated for triggers and uploaded.
\section{Derivation of $P(t_k,\rho_k|t_j,\rho_k)$}
\label{appendix:tw1}
In this appendix we derive Eq.~\eqref{eq:confusion} which evaluates the probability of a template $t_k$ being triggered by a GW signal that matches exactly with template $t_j$. The GW strain data in a detector can be represented by a time-series $d$ which can be expressed as a superposition of a potential signal $s$ and noise $n$ in the following way:
\begin{equation}
    d=s+n
\end{equation}.
Matched filtering involves computing the noise averaged inner product of this time-series data with the time-series corresponding to a template waveform. If the true signal is a perfect match to the jth template then the matched filter corresponding to the kth  template can be expressed in the following way:

\begin{equation}
\rho_k =\rho_jt_j\cdot t_k   +n_k
\end{equation}
,where $\rho_k =t_k\cdot d$ is the matched filter SNR corresponding to template $t_k$, $n_k=t_k \cdot n$ and $\rho_j$ is the SNR of the true signal $s$. As mentioned in Sec.~\ref{methods:LR}, we are interested in quantifying how noise fluctuations might cause a true signal $t_j$ to be highest \ac{SNR} match with template $t_k$. Hence, as a first approximation, we can set $\rho_j=\rho_k$. Under this approximation and the assumption that the noise is stationary and Gaussian, we can compute the probability of template migration:

\begin{equation}
    P(t_k,\rho_k|t_j,\rho_j)\propto\frac{1}{\sqrt{2\pi}} e^{-\frac{1}{2}|n_k|^2}=\frac{1}{\sqrt{2\pi}}e^{-\frac{1}{2}\rho_k^2\left[1-2t_k\cdot t_j+(t_j\cdot t_k)^2\right]}
\end{equation}
\section{Template weights $P(t_k,\rho_k|H_a)$: variation with template parameters and SNR}
\label{appendix:tw2}
\begin{figure*}[htt]
\centering
\includegraphics[width=0.95\textwidth]{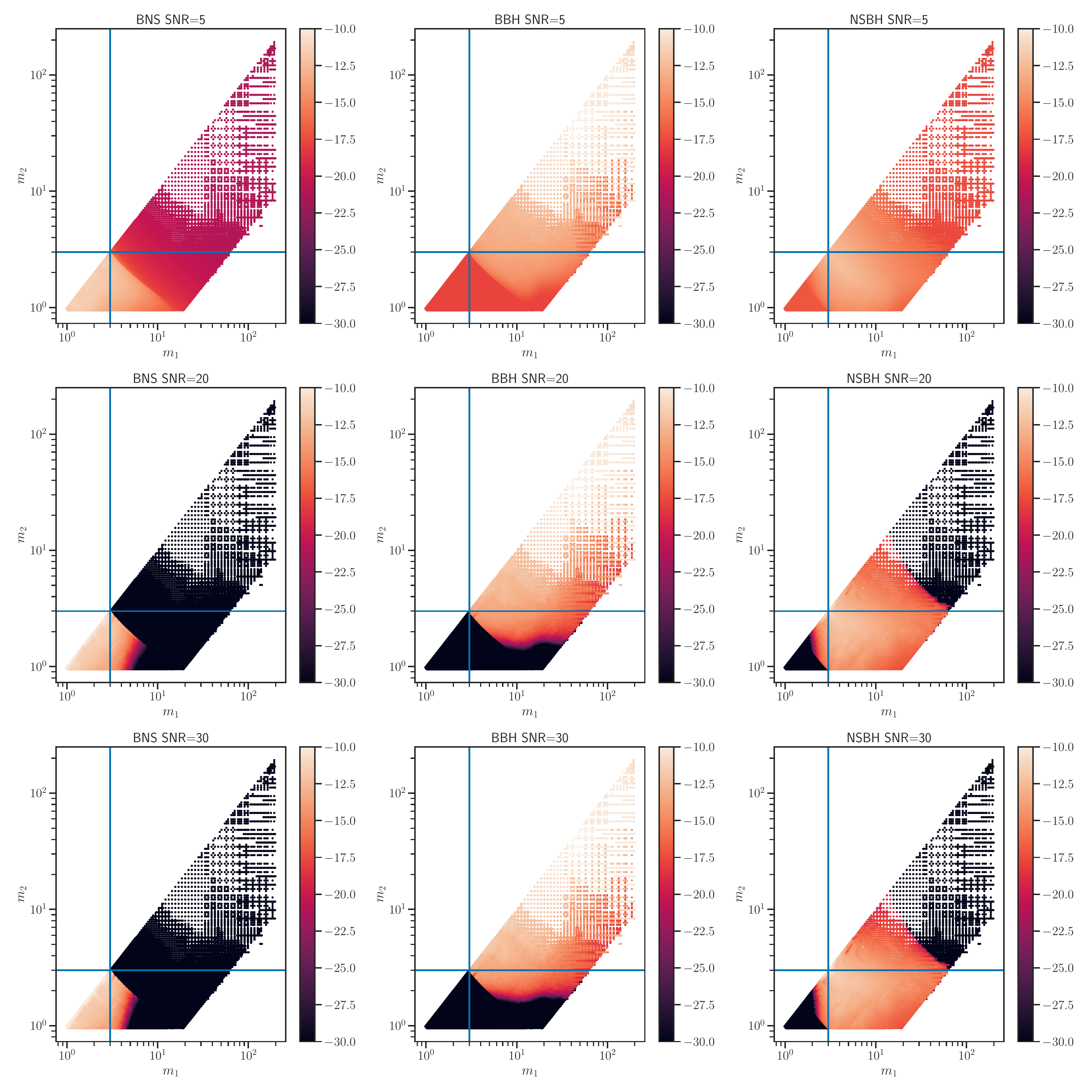}
\caption{\label{fig:template-weights-salpeter-m1m2} Variation of the template weights corresponding to the Salpeter population distribution~(i.e. the ones to be used in O4), as a function of component masses and SNR. The colorbars represent $\log P(t_k,\rho_k|H_a)$ with $a$ and $\rho$ specified at the top of each panel.}
\end{figure*}
In this appendix, we summarize how the template weights vary as a function of the template parameters and SNR. Given the form of the migration probability in Eq.~\eqref{eq:confusion}, we expect there to be significant confusion among source categories that map onto neighboring regions of the template-bank at low SNR. As the SNR increases, we expect the probability of miss-classification to decrease with perfect classification at infinite SNR. 

\begin{figure*}[htt]
\centering
\includegraphics[width=0.95\textwidth]{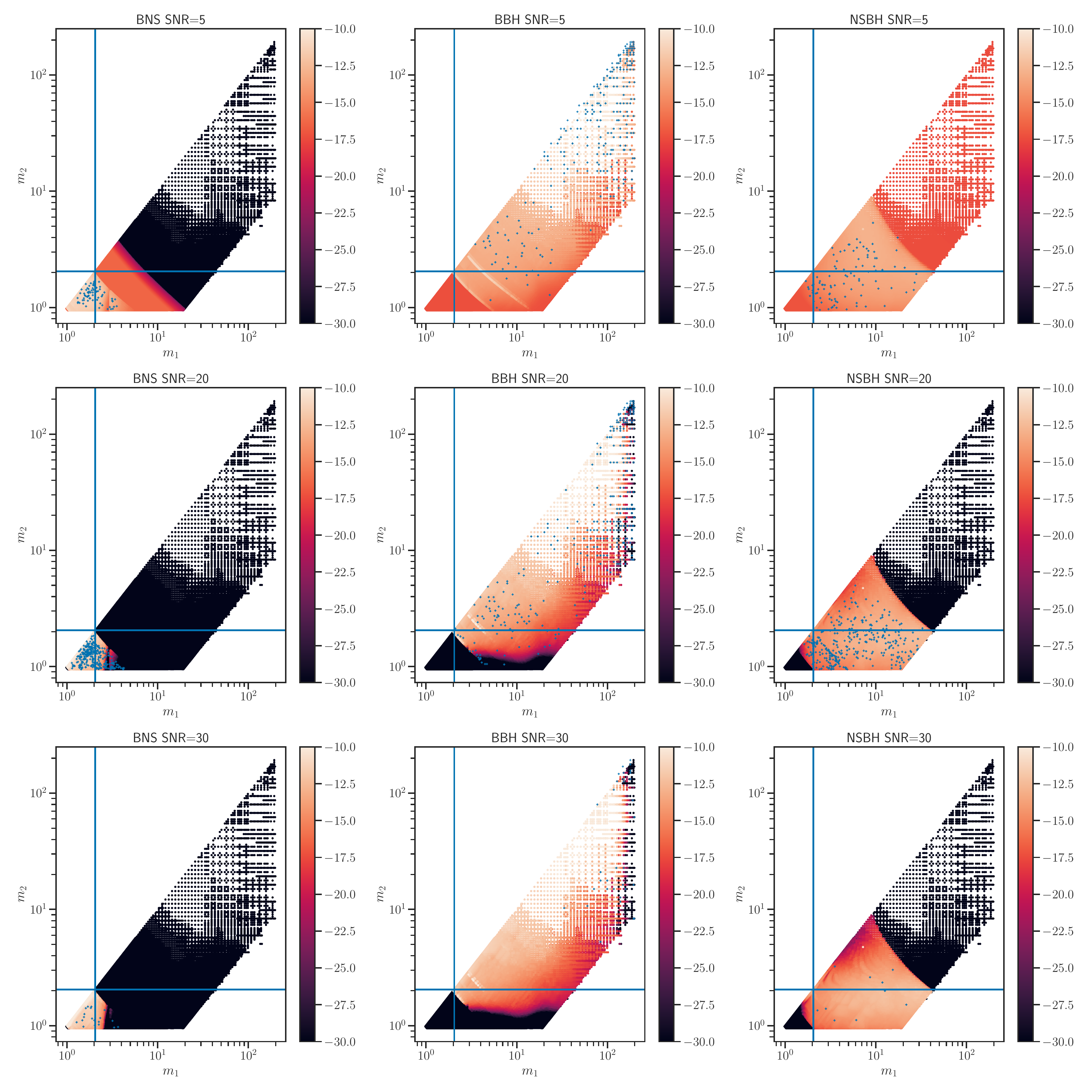}
\caption{\label{fig:template-weights-inj-m1m2} Variation of the template weights corresponding to the population distribution of injections, as a function of component masses and SNR. The colorbars represent $\log P(t_k,\rho_k|H_a)$ with $a$ and $\rho$ specified at the top of each panel. The blue dots in each panel correspond to recovered masses of injections whose true source category coincides with the one specified on top of the panel. The \acp{SNR} of the injections in each panel lie within a narrow range about the \ac{SNR} specified on top of the pannel}
\end{figure*}

For the Salpeter function mass-model in Eq.~\eqref{astromass-model} that shall be used by \GSTLAL{} to compute \PASTRO{} values in low-latency for O4, the template weights as a function of template masses are displayed in Fig.~\ref{fig:template-weights-salpeter-m1m2}. We find that the variation with SNR is exactly what we expect them to be.

Similarly, for the population model corresponding to the MDC injections (summarized in Eq.~\eqref{injmass-model}), the template weights as a function of component masses is displayed in Fig.~\ref{fig:template-weights-inj-m1m2}  We also over plot the recovered parameters of the injections from category $a$ found by the pipeline above threshold with SNR $\rho$, atop the plots of the template weights $P(t_k,\rho|H_a)$.

It can be seen that the density of blue points at various regions matches with the heat map representing $\log{P(t_k,\rho|H_a)}$. For example, in some \ac{SNR} ranges wherein \ac{BNS} injections are recovered in the \ac{NSBH} region, $P(t_k,\rho|H_{\ac{BNS}})$ in that region corresponding to the \ac{SNR} in question is also found to have non-zero value. This is implicative of the fact our template weights correctly model the mismatch between the recovered template and the true signal due to noise fluctuations, provided the population distribution corresponding to each category is exactly known.

\clearpage

\printbibliography
\end{document}